\documentclass[aps,showpacs,sort&compress,footinbib]{revtex4}%
\usepackage{amssymb}
\usepackage{amsmath}
\usepackage{amscd}
\usepackage{graphicx}
\usepackage{subfigure}
\usepackage{float}

\begin{document}

\title{Arbitrary Control of Entanglement between two nitrogen-vacancy center ensembles coupling to superconducting circuit qubit}
\author{Wan-Jun Su}
\email{wanjunsu@fzu.edu.cn}
\affiliation{Department of Physics, Fuzhou University, Fuzhou
350002, People's Republic of China}
\affiliation{ Institute for Quantum Information Science, University of Calgary, Alberta T2N 1N4, Canada}
\author{Zhen-Biao Yang}
\author{Zhi-Rong Zhong}
\affiliation{Department of Physics, Fuzhou University, Fuzhou
350002, People's Republic of China}


\date{\today}

\begin{abstract}
We propose an effective scheme for realizing a Jaynes-Cummings (J-C) model with the collective nitrogen-vacancy center ensembles (NVE) bosonic modes in a hybrid system. Specifically, the controllable transmon qubit can alternatively interact with one of the two NVEs, which results in the production of $N$ particle entangled states. Arbitrary $N$ particle entangled states, NOON states, N-dimensional entangled states and entangled coherent states are demonstrated. Realistic imperfections and decoherence effects are analyzed via numerical simulation. Since no cavity photons or excited levels of the NV center are populated during the whole process, our scheme is insensitive to cavity decay and the spin dephasing effect of NVE. The idea provides a scalable way to realize NVEs-circuit cavity quantum information processing with current technology.
\end{abstract}

\pacs{03.65.Xp,03.65.Vf,42.50.Dv,42.50.Pq}

\maketitle

\section{Introduction }

Recently, with the advantages of the cavity or circuit QED systems \cite{RaimondRMP01,MakhlinRMP01,WallraffNature04,BlaisPRA04,MariantoniNP11,PRL015502}, the hybrid systems have attracted much attention for quantum information applications. For example, the composite system consisting of a NVE, a superconducting microstrip (SCM) cavity and a transmon qubit, has emerged as one of excellent candidates for the construction of solid-state quantum information processor \cite{YouNature11, APRL09,PRAPPl044003}. In the low-excitation limit, the collective excitations of a NVE behave as bosonic modes. What's more, the collectively enhanced magnetic dipole interaction will result in obtaining appreciable coupling strength, which has been proven in recent experiments \cite{KuboPRL10,RPRL11,SchusterPRL10}.

The question of creating an arbitrary quantum state of a cavity field
or atoms (ion, NV centers) has been discussed in these papers
\cite{LawPRL96,StrauchPRL10,YangPRA12,PRA032342,PRA042306,SadowskiarXiv:1303.3757}.
The method proposed by Law and Eberly \cite{LawPRL96} was using a two-channel approach. In their scheme, a controllable coupling strength between atom-cavity and a driving classical field with an adjustable amplitude have been used. Based on Law and Eberly's work, Strauch et al \cite{StrauchPRL10} have presented a method to synthesize an arbitrary quantum state of two superconducting resonators using a control qubit. They used the photon-number-dependent Stark shift to achieve selectively manipulations of the quantum system. Yang et. al. \cite{YangPRA12} have presented a scheme to engineer a two-mode squeezed state of effective bosonic modes. The collective excitations of two distant NVEs were coupled to separated transmission line resonators (TLRs). By engineering NVE-TLR magnetic coupling with Raman transition between the ground levels of the NVEs, they may manipulate the artificial reservoir by tuning the external driving fields. Recently, Li et. al. \cite{PRA032342} have proposed a scheme for a coherent quantum microwave-optical interface mediated by an NV center ensemble. Quantum state conversion can realized using the collective spin excitation modes. Up to now, Fock states\cite{BertetPRL02,HofheinzNature08}, Schr$\ddot{o}$dinger cat state \cite{MeunierPRL05} and entangled coherent states \cite{RauschenbeutelSCi00} have been produced in cavity or circuit QED experiments.

Motivated by these works above, we propose a scheme to engineer arbitrary
entangled states of two distant NVEs coupled to a SCM with a tunable quantum qubit.
Meanwhile, based on the same model, multi-dimensional entangled state, NOON state and entangled coherent states of NVEs are also produced. In this scheme, we add a
driving microwave or magnetic field with an adjustable amplitude \cite{LawPRL96}, to control a suitable dispersive interaction. Treating the collective NVEs spin as a bosonic mode, we set up an effective J-C model \cite{SPRA08}. What's more, a resonant J-C model can be switched to a non-resonant J-C model by tuning the Rabi frequency of the microwave or magnetic field. In contrast to the previous schemes, the present approach has the following merits: (i) We can realize arbitrary control of entanglement between two NVEs by
selective manipulation of the control qubit. (ii) The cavity field would not be excited during the whole process because the interaction is a virtual-photon process. In other words, our model works well in the bad-cavity limit, which makes it more applicable to current laboratory techniques. (iii) More importantly, our idea can be generalized to generate entangled states for two or more NVEs, providing a potentially practical tool for large-scale one-way quantum computation.

\section{Physical model and effective dynamics}

\begin{figure}[tbp]
 \centering
  \includegraphics[width=0.45\textwidth]{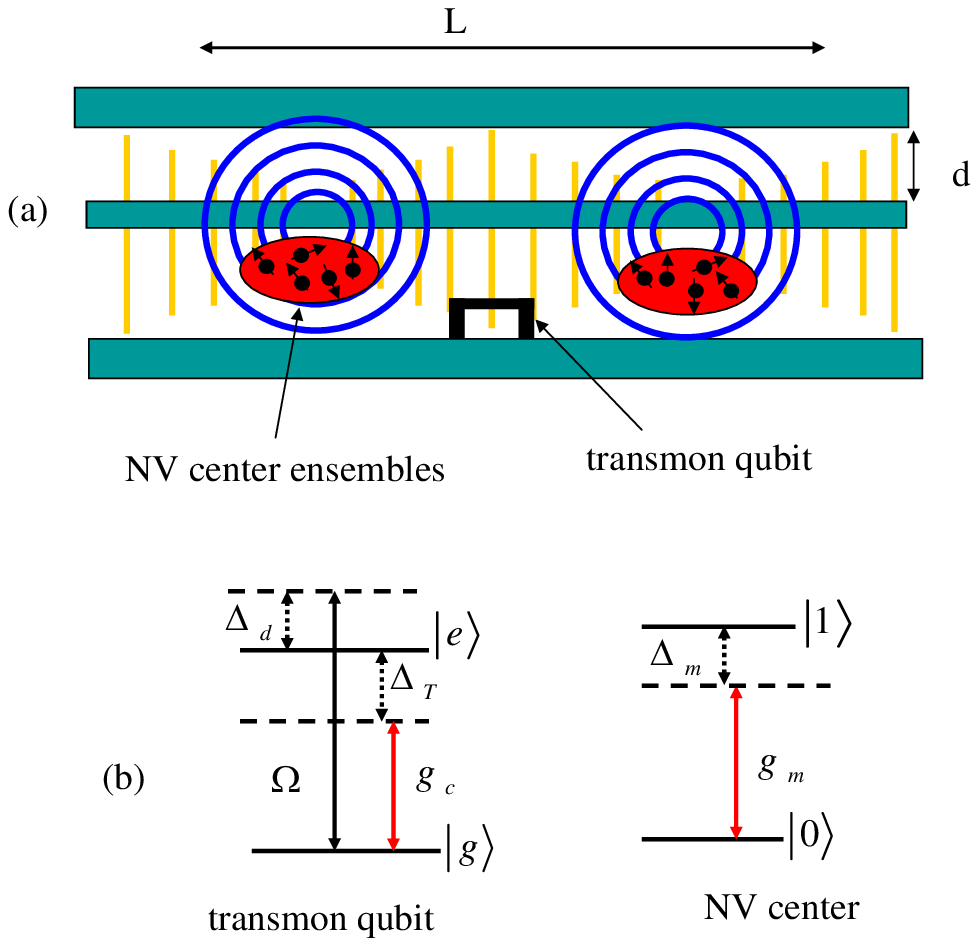}
  \caption{(Color online) (a) The experimental setup for arbitrary controlling of entanglement between two NVEs coupling to superconducting circuit qubit in a SCM cavity with $d\sim10\mu m$ and $L\sim 1cm$. The length of the orange lines indicates the strength of the cavity-mode electric field. The blue lines that encircle the center conductor depict the magnetic field lines at the locations where their strength is maximum. The presence of a transmon qubit at an electric-field maximum ensures that the cavity has a large nonlinearity. The NVE of a diamond crystal trapped $\sim10 \mu m$ above the cavity structure. (b) Configurations of the transmon qubit and the NV level structure and relevant transitions.}
\end{figure}
The basic idea of this work is illustrated by the schematic setup
shown in Fig. 1. Two NVEs (denoted 1 and 2) with different frequencies
are coupled to a tunable superconducting circuit qubit in a SCM cavity.
The SCM cavity with frequency $\omega_c$, serves as a quantum bus, which is electrostatically coupled to the control qubit and
magnetically coupled to the NVEs. The NVE consists of
$\sim 10^{12}$ negatively changed NV color centers in a diamond
crystal. The ground state of NV center is a spin triplet, which
labeled as $^3A$. There is a zero-field splitting (2.88GHz) between
the state $\left| 0\right\rangle $ ($m_s=0$) and $\left| \pm
1\right\rangle $($m_s=\pm1$) for the spin-spin interaction \cite{LloydSCi93}.  We
consider the ground state $|m_s=0\rangle=|0\rangle_{NV}$ and the
excited state $|m_s=\pm 1\rangle=|1\rangle_{NV}$, with the corresponding transition
energy $\hbar\omega_{NV}$. The tunable qubit is a transmon qubit, with the
ground state $|g\rangle$ and excited state $|e\rangle$, which are separated by level energy $\hbar\omega_T$. The qubit is coupled to a nonlinear resonator which is used to read out its state or apply a driving microwave, as in related circuit QED experiments
\cite{MalletNP09}. Through the nonlinear resonator, the qubit is driven
by a microwave field with Rabi frequency $\Omega_{(t)}$ and frequency
$\omega_{d}$. The transmon qubit is introduced at
an electric field maximum of the superconducting cavity. In
contrast, each of the NVEs is separately placed at the corresponding locations where
the magnetic field is maximum. As the collective enhanced couplings are employed, here we introduce $g_{m}=\sqrt{\frac{1}{N}\sum_{i=1}^{N}|g_{m}^{i}(r_{i})|^{2}}$ to denote the average magnetic dipole coupling strength for each spin to the cavity, and the collective spin coupling strength $g_{eff}=\sqrt{N}g_{m}$, where $N$ is the number of spins \cite{PRL220501}. The presence of random local strain may inhomogeneously broaden the transition frequencies. The corresponding random shifts are $\delta_{m}^{i}=\Delta_{m}^{i}-\Delta_{m}$, where $ \Delta_{m}=\omega_{NV}-\omega_c$ is the average detunings. In the frame rotating with the cavity frequency $\omega_c$, and considering the detunings for the related transitions $\Delta_{T}=\omega_T-\omega_c$, $\Delta_{d}=\omega_T-\omega_d$, the Hamiltonian of the combined system is given by ($\hbar=1$)

\begin{eqnarray}\label{1}
H&=&[g_c\sigma_{eg}a_{c}e^{i\Delta_{T}t}+\Omega_{(t)}\sigma_{eg}e^{i\Delta_{d}t} \\&&
+\sum_{i=1}^{N}\sum_{j=1,2}g_{m_{j}}\sigma_{i_{j}}^{+}a_{c}e^{i\Delta_{m_{j}}t}]+H.c.
\end{eqnarray}
where $j=1,2$ denotes NVE 1 or 2. $\sigma_{eg}=|e\rangle\langle g|$ and
$\sigma_{i_{j}}^{+}=|1\rangle_{NV_{j}}\langle 0|$ denote the raising
operator of the transmon qubit (NV center). $a_c$ is the annihilation operator of the superconducting cavity mode. $g_c$ denotes the electric dipole coupling strength of the transmon qubit to the cavity mode. Under the condition $|\Delta_{T}|, |\Delta_{d}|\gg|g_c|,|\Omega_{(t)}|$, the classical field and the cavity field induce Stark shift. In the case $|\Delta_{m_{j}}|\gg| g_{m_{j}}\sqrt{N}|$, the dispersive interaction between NVEs and cavity field leads to the Stark shift and dipole coupling for the NV centers. For the large dunting, we can ignore the inhomogeneous broadening of the transition frequencies in the following. If the cavity field initially is in the vacuum state, the photon number will remain zero during the whole process. The Hamiltonian can be rewritten as \cite{ZhengPRL00},
\begin{eqnarray}\label{2}
H&=&(\frac{2\Omega_{(t)}^2}{\Delta_{d}}+\frac{g_{c}^2}{\Delta_{T}})S_z
+\sum_{i,k=1}^{N}\sum_{j=1,2}\frac{g_{m_{j}}^2}{\Delta_{m_{j}}}\sigma_{i_{j}}^{+}\sigma_{k_{j}}^{-}\nonumber\\&&
+\sum_{i=1}^{N}\sum_{j=1,2}\frac{g_{m_{j}}g_{c}}{2}(\frac{1}{\Delta_{m_{j}}}+\frac{1}{\Delta_{T}})
(\sigma_{i_{j}}^{+}\sigma_{ge}e^{-i\xi_{j}
t}+\sigma_{i_{j}}^{-}\sigma_{eg}e^{i\xi_{j} t}),\nonumber\\&&
\end{eqnarray}
where $S_z=(|e\rangle\langle e|-|g\rangle\langle g|)/2$ and the constant energy $({g_{c}^2}/{2\Delta_{T}})(|e\rangle\langle e|+|g\rangle\langle g|)$ has been
discarded. Introducing a collective operator $b^{+}=1/\sqrt{N}
\sum_{i=1}^{N}\sigma_{i}^{+}$, in the low excitation limit, where almost all NV centers are in the ground state, the operator $b^{+}$ and $b$ obey approximately bosonic commutation relations $[b,b^{+}]=1$. In this case $\sum_{i,j=1}^{N}\sigma_{i}^{+}\sigma_{j}^{-}=Nb^{+}b$, so a NVE can
be considered as an effective bosonic mode. Considering $\Delta_{T}=\Delta_{d}=\Delta$, the Eq. (2) can be reduced to an effective J-C model, which denotes the interaction between a two-level qubit and two bosonic modes. The effective Hamiltonian is given by
\begin{eqnarray}\label{3}
H_{eff}&=&\omega_{z}(t)S_z+\omega_{b_{1}}b_{1}^{+}b_{1}+\omega_{b_{2}}b_{2}^{+}b_{2}\nonumber\\&&
+G_{1}(\sigma_{ge}b_{1}^{+}+\sigma_{eg}b_{1})+G_{2}(\sigma_{ge}b_{2}^{+}+\sigma_{eg}b_{2}).
\end{eqnarray}
where $\omega_z(t)={2\Omega_{(t)}^2}/\Delta+{g_{c}^2}/\Delta$ and
$\omega_{b_{j}}={g_{m_{j}}^{2}N}/\Delta_{m_{j}}$. $G_{j}=\sqrt{N}g_{m_{j}}g_{c}/\Delta_{m_{j}}$ corresponds to the effective coupling strength. Changing the detunings $\Delta_{m_{j}}$ and the Rabi frequency $\Omega$, we can dynamically control $\omega_z(t)$. These effective couplings $G_{j}$ and the scaled frequency $\omega_{b_{j}}$ can be dynamically controlled by the detunings $\Delta_{m_{j}}$.

\begin{figure}[tbp]
 \centering
  \includegraphics[width=0.5\textwidth]{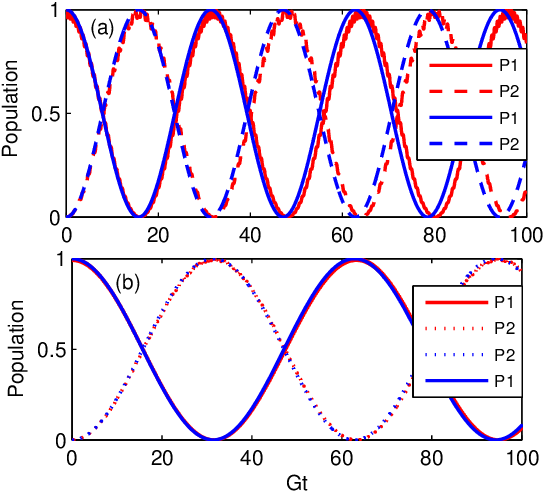} \caption{(Color online)
The populations of the basic states
$|e\rangle|0_{1}\rangle|0_{2}\rangle|0\rangle_c$ (P1) and
$|g\rangle|1_{1}\rangle|0_{2}\rangle|0\rangle_c$ (P2) governed by the full
Hamiltonian in Eq. (1) (red lines) and the effective Hamiltonian in
Eq. (3) (blue lines). Parameters: $G_{1}=\sqrt{N}g_{m_{1}}g_{c}/\Delta=G$,  $G_{2}=0$, $g_{m_{1}}=g_{c}/\sqrt{N}$, $g_c=10G$, $\Omega=G$,
$N\sim10^{12}$, (a) $\Delta=100G$, (b) $\Delta = 200G$.}
\end{figure}

In order to validate the feasibility of the above physical model, we perform a direct numerical simulation the Schr\"{o}dinger equation with the full Hamiltonian and the effective Hamiltonian. For simplicity, the interaction of one NVE, a transmon qubit, and a cavity is considered. Setting the parameters $G_{1}=G$, $G_{2}=0$, $g_{m_{1}}=g_{c}/\sqrt{N}$, $g_c=10G$, $\Omega=G$, $N\sim10^{12}$ and $\Delta=100G$, we can satisfy the resonant condition  $\omega_{z}(t)\approx\omega_{b_{1}}=G$. In the following simulation, we calculate the temporal evolution of the system with the initial state $|e\rangle|0\rangle_{NVE}|0\rangle_c$. We plot the time-dependent populations $\rho(t)=|\psi(t)\rangle\langle\psi(t)|$ of the basic states $|e\rangle|0\rangle_{NVE1}|0\rangle_{NVE2}|0\rangle_c$ (P1) and $|g\rangle|1\rangle_{NVE1}|0\rangle_{NVE2 }|0\rangle_c$ (P2) governed by the full Hamiltonian (red lines) and the effective Hamiltonian (blue lines). Fig. 2 shows that the effective and full dynamics exhibit excellent agreement whether the $\Delta$ is equal to $100G$ or $200G$, and the larger $\Delta$ gets a better result. However, the deviation decrease is at the cost of the long evolution time. Eventually, the simulation result of full Hamiltonian is almost the same as that of the effective Hamiltonian when $\Delta = 200G$. Thus, the above approximation for the Hamiltonian is reliable as long as $\Delta$ is large enough.

Choosing the microwave pulse to satisfy $\omega_{z}(t)=\omega_{b_{j}}$ (j=1 or 2), which the qubit frequency is between two operating points, the control qubit is resonantly coupling with one NVEs bosonic mode (turn on the coupling $G_{1}$), meanwhile well off-resonance with the other one ($G_{2}=0$). In this case, the transition is given by
\begin{eqnarray}\label{4}
|e\rangle|n_{1}\rangle|n_{2}\rangle&\rightarrow&
e^{-i\omega_{b_{1}}t}[\cos(G_{1}\sqrt{n_{1}+1}t)|e\rangle|n_{1}\rangle -i\sin(G_{1}\sqrt{n_{1}+1}t)|g\rangle|n_{1}+1\rangle]|n_{2}\rangle,\cr
|g\rangle|n_{1}+1\rangle|n_{2}\rangle&\rightarrow&
e^{-i\omega_{b_{1}}t}[\cos(G_{1}\sqrt{n_{1}+1}t)|g\rangle|n_{1}+1\rangle
-i\sin(G_{1}\sqrt{n_{1}+1}t)|e\rangle|n_{1}\rangle]|n_{2}\rangle,
\end{eqnarray}
where $|n\rangle$ denotes the Fock state for the collective NVEs
spin. The system will oscillate between $|e\rangle|n_{1}\rangle|n_{2}\rangle$ and the
state $|g\rangle|n_{1}+1\rangle|n_{2}\rangle$ at an angular frequency  $G_{1}\sqrt{n_{1}+1}$. The dynamics provides a way to realizing various interesting
phenomena in this hybrid system.

\begin{figure}[tbp]
 \centering
  \includegraphics[width=0.5\textwidth]{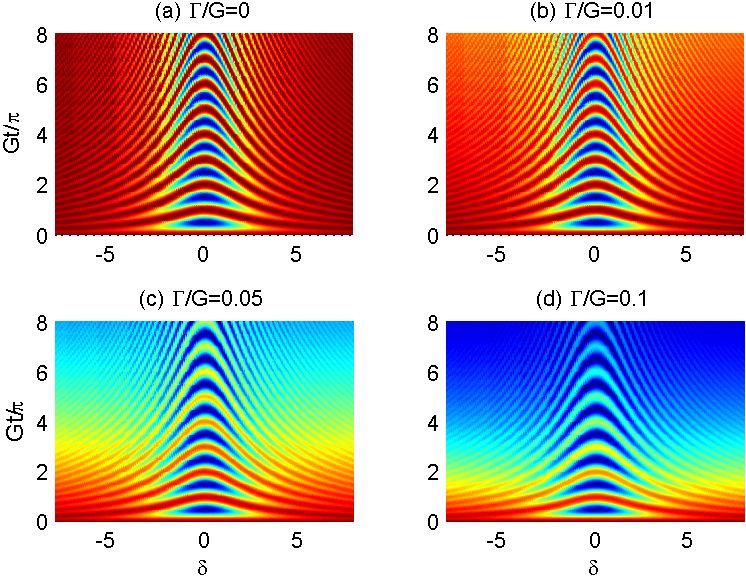} \caption{(Color online)
The swap spectrum between qubit and one NVE in scaled detuning-interaction time plane. The qubit detuning can be adjusted \cite{MalletNP09}, here, $\delta=(\omega_{z}(t)-\omega_{b_{1}})/G$, $G_{1}=G$, $G_{2}=0$. Considering the cases in according to the qubit different scaled relaxation rate $\Gamma/G$, (a) $\Gamma/G=0$, (b) $\Gamma/G=0.01$, (c) $\Gamma/G=0.05$, (d) $\Gamma/G=0.1$.}
\end{figure}

To control this system, we modify the time-dependent qubit frequency $\omega_{z}(t)$. For instance, changing the frequency and intensity of the shift pulse causes the NVEs and the qubit to be tuned in and out of resonance with each other. We can pump photons one at a time into the NVE by repeatedly exciting the detuned qubit from $|g\rangle$ to $|e\rangle$ using a qubit microwave $\pi$-pulse, followed by a controlled-time, on-resonance photon swap. The swap spectrum \cite{MarkkuPRL14} between qubit and one NVE in scaled detuning-interaction time plane is shown in Fig. 3. The qubit detuning is $\delta=(\omega_{z}(t)-\omega_{b_{1}})/G$. Here, we consider the qubit different scaled relaxation rate $\Gamma/G$. As a result, the effective coupling between NVEs and the qubit may be turned on and off. By performing a sequence of microwave pulses, the transmon qubit alternately resonantly interacts with two NVEs. Quanta can be created and transferred between the qubit and the two NVEs.

\section{arbitrary entangled states}

A corresponding algorithm for the synthesis of an arbitrary quantum state of two NVEs is described as
\begin{eqnarray}\label{5}
|\psi\rangle=\sum_{n_1=0}^{N_1}\sum_{n_2=0}^{N_2}c_{n_1,n_2}|n_1,n_2\rangle.
\end{eqnarray}
where $|n_{1},n_{2}\rangle$ is Fock states of collective bosonic mode of two NVEs.

Now we use the physics model above to generate arbitrary entangled states for two NVEs $|n_{1},n_{2}\rangle$, whereas this mechanism applying in the synthesize an arbitrary quantum state of two superconducting resonators \cite{StrauchPRL10}. For definiteness, the system state denotes $|q,n_{1},n_{2}\rangle$ (where the qubit state is q=e or g). The resonant interaction of the qubit with each NVE leads to creating Fock states based on the $H_{eff}$ in Eq. (3). These resonant interactions are efficient and fast so that the interaction time is so short to prevent the decoherence. However, in order to generate an arbitrary quantum state of two NVEs, an independent state-selective qubit rotation is required. How to address the qubit between the resonant interactions is an important problem. To answer this question, a driving microwave pulse (Rabi frequency $\Omega_{s}$) can be used to make the photon-number-dependent Stark shift take effect \cite{FPRL11,AlexandrePRA04}. This implies that in the dispersive regime,the transmon qubit will undergo Rabi oscillations from $|g,n_{b_1},n_{b_2}\rangle\rightarrow |e,n_{b_1},n_{b_2}\rangle$. To make this process happen, the frequency of the driving microwave satisfies
\begin{eqnarray}\label{6}
\omega_s=\omega_z(t)+\frac{{G_1}^2}{\delta_1}(2n_{b_1}+1)+\frac{{G_2}^2}{\delta_2}(2n_{b_2}+1),
\end{eqnarray}
where $\delta_{1}=\omega_{z}(t)-\omega_{b_1}$ and $\delta_{2}=\omega_{z}(t)-\omega_{b_2}$. We set $\lambda={G_{1}}^2/\delta_1=-{G_{2}}^2/\delta_2$, which can always be achieved for the transmon qubit with a tunable frequency $\omega_z(t)$ by applying the "shift" pulses $\omega_d$.  To avoid nonresonant transitions, we assume $|\Omega_s|<\lambda$ and  $\omega_{b1}<\omega_{z}(t)<\omega_{b2}$. So we can simplify the frequencies of the driving microwave
\begin{eqnarray}\label{7}
\omega_n=\omega_z(t)+2\lambda \Delta n,
\end{eqnarray}
where $\Delta n=n_{b_1}-n_{b_2}$ is an integer. By choosing different values of $\omega_n$, we can address each of the qubits between the resonant interactions.

\begin{figure}[tbp]
 \centering
  \includegraphics[width=0.45\textwidth]{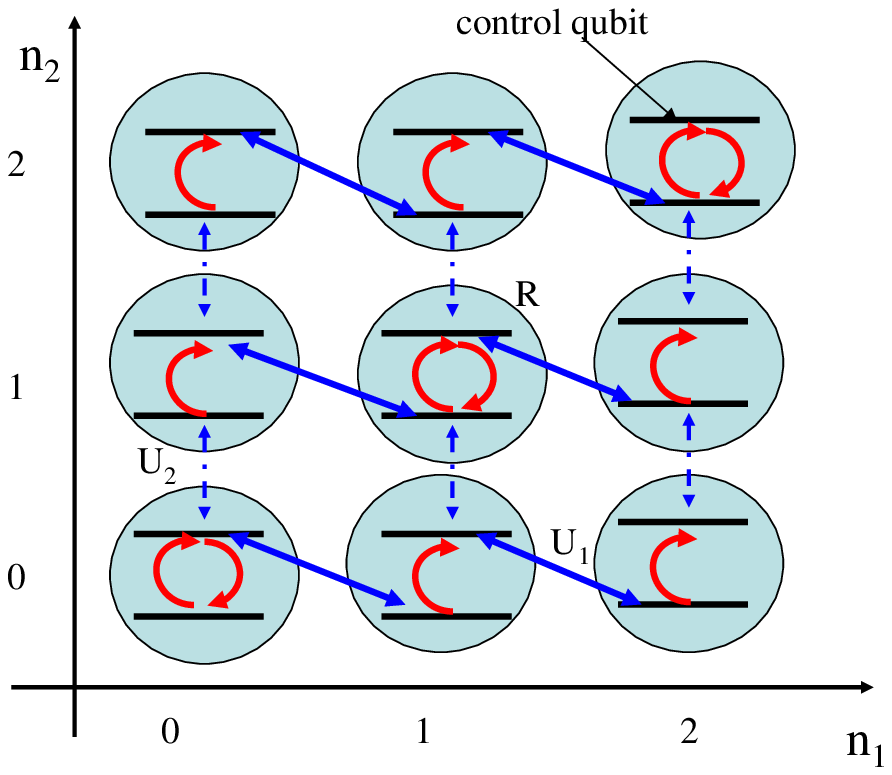} \caption{(Color online)
Schematic set of operations to generate an arbitrary state of two NV
center ensembles. In this Fock-state diagram, the state
$|q, n_1, n_2\rangle $ is represented by the node at location ($n_1,
n_2$), where q is control qubit state( q=e or g). Interactions lead to
couplings between these states, indicated by the arrows. Three key
interactions are used: $U_1$ transfers quanta between the qubit and
NVEs 1 (solid lines), $U_2$ transfers quanta between the qubit and
NVEs 2 (dashed lines), and $R $ (curved arrows) rotates the qubit. }
\end{figure}

Assuming that the control qubit is in the ground state and the collective modes of NVEs are in the vacuum states, the initial state of the system is

\begin{eqnarray}\label{8}
|\psi_0\rangle=|g,0,0\rangle.
\end{eqnarray}
Our goal is to force the system to evolve into a final state of the form
\begin{eqnarray}\label{9}
|\psi_{(t)}\rangle=\sum_{n_1=0}^{N_1}\sum_{n_2=0}^{N_2}c_{n_1,n_2}|g,n_1,n_2\rangle.
\end{eqnarray}

The time evolution operator of the system can be expressed as a product of evolution operators associated with the time intervals, which is accomplished by the following sequence of operations:
\begin{eqnarray}\label{10}
U(t)=[\prod_{j=1}^{N_2}(\prod_{k=0}^{N_1}U_{2,jk}R_{2,jk})]\prod_{j=1}^{N_1}U_{1,j}R_{1,j},
\end{eqnarray}
where $U_{1,j}$ and  $U_{2,j}$ describe the evolution due to resonantly interaction between the NVEs and the qubit, corresponding to Eq. (4). The microwave $\Omega_s$ turns off when $U_{1,j}$ and $U_{2,j}$ do work. $R_{1,j}$ and $R_{2,j}$ describe the evolution due to the single-qubit rotations, which use the Stark-shifted Rabi pulses. $R|\psi_{(t)}\rangle=\sum_{n_1=0}^{N_1}\sum_{n_2=0}^{N_2}r_{q,n_1,n_2}|g,n_1,n_2\rangle$.
If the collective bosonic modes in different NVEs satisfy
$n_1-n_2=\Delta n$, the factors $r_{q,n_1,n_2}$ are
\begin{eqnarray}\label{11}
r_{e,n_1,n_2}&=&e^{-i\alpha}\cos(\Omega_s
t)c_{e,n_1,n_2}-ie^{-i\beta}\sin(\Omega_s t)c_{g,n_1,n_2}\cr
r_{g,n_1,n_2}&=&e^{i\alpha}\cos(\Omega_s
t)c_{g,n_1,n_2}-ie^{i\beta}\sin(\Omega_s
t)c_{e,n_1,n_2}.
\end{eqnarray}
While $n_1-n_2\neq \Delta n$, $r_{q,n_1,n_2}=e^{i\phi_q}c_{q,n_1,n_2}$.

To determine the precise sequence of operations for a given state $|\psi_{(t)}\rangle$, one solves the equation of inverse evolution,
\begin{eqnarray}\label{12}
|g,0,0\rangle&=&U_{(t)}^{\dag}|\psi_{(t)}\rangle\nonumber\\&=&
\prod_{j=1}^{N_1}R_{1,j}^{\dag}U_{1,j}^{\dag}[\prod_{j=1}^{N_2}
(\prod_{k=0}^{N_1}R_{2,jk}^{\dag})U_{2,jk}^{\dag}]|\psi_{(t)}\rangle.
\end{eqnarray}
Each step of the sequence in the right side of Eq. (12) can remove bosonic exciton successively from the state $|\psi_{(t)}\rangle$ until all the excitons are exhausted. In the Fock-state diagram as shown in Fig. 4, $R_{2,jk}^{\dag}U_{2,jk}^{\dag}$ acts on $|\psi_{(t)}\rangle$, and moves the system along the vertical paths. All populations in $|g,n_k,n_{j}\rangle$ and $|e,n_k,n_{j-1}\rangle$ are transferred to $|g,n_k,n_{j-1}\rangle$. After all the excitons have been removed from the columns in row $j$, the sequence repeats for $j-1$. The step will not stop until the exciton of NVE 2
bosonic mode is zero, and then the system state is $|g,n_k,0\rangle$. Once there, the $R_{1,j}^{\dag}U_{1,j}^{\dag}$ sequence moves population along the horizontal paths to
$|g,0,0\rangle$. By counting the number of operations in $U$, we find that the general sequence requires $N_1$ $U_{1,j}^{\dag}$ unitary, $(N_1+1)N_2$ $U_{2,j}^{\dag}$ unitary, and about $(N_1+1)(N_2+1)$ Rabi pulses. So the total interaction time is approximately given by
\begin{eqnarray}\label{13}
t_{max}=(N_1+1)(N_2+1)\frac{\pi}{\Omega_{s}}+
\frac{\pi}{G_{1}}\sum_{j=1}^{N_1}\frac{1}{\sqrt{j}}+(N_1+1)\frac{\pi}{G_{2}}\sum_{j=1}^{N_2}\frac{1}{\sqrt{j}}
\end{eqnarray}

\begin{figure}[tbp]
 \centering
 \subfigure
  {\includegraphics[width=0.5\textwidth]{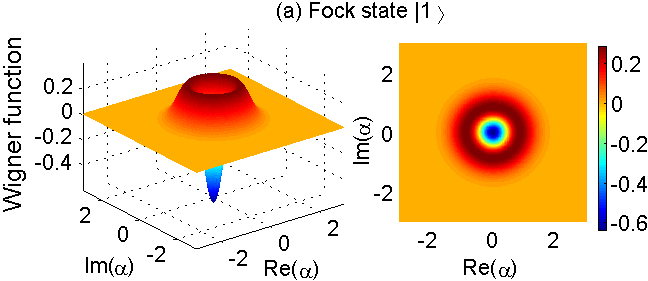}}
  \subfigure
  {\includegraphics[width=0.5\textwidth]{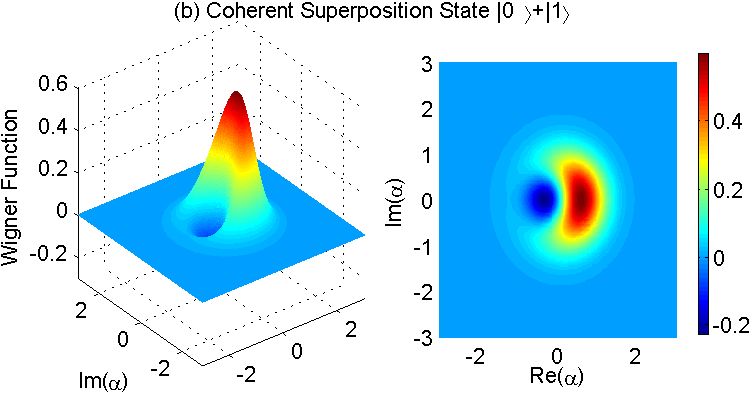}}
  \subfigure
  {\includegraphics[width=0.5\textwidth]{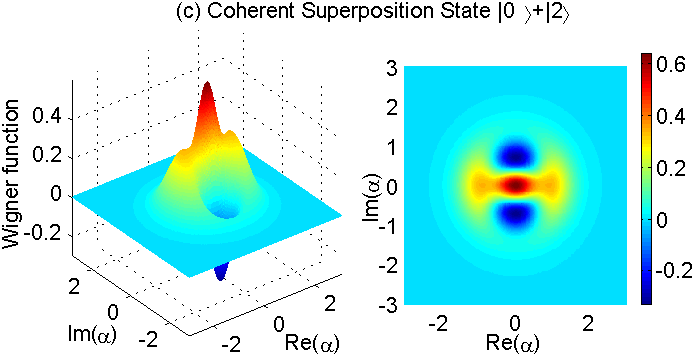}}
  \caption{(Color online) The Wigner function encodes information about the
amplitudes and coherences of quantum states. For example, we choose (a) the Fock state $|1\rangle$, (b) the coherent superposition states $|0\rangle+|1\rangle$, (c) the coherent superposition states $|0\rangle+|2\rangle$. }
\end{figure}
After the preparation, we analyse the NVEs states using Wigner tomography \cite{WangPRL09}. The Wigner tomography can map out the Wigner quasiprobability distribution $W(\alpha)$ as a function of the phase space amplitude $\alpha$ of the NVEs. The Wigner function $W(\alpha)$  and density matrix $\rho$ are related via the trace
\begin{eqnarray}\label{14}
W(\alpha)=\frac{2}{\pi}Tr[\rho D(\alpha)e^{i\pi a^{+}a}D(-\alpha)]
\end{eqnarray}
For simplicity, here, we assume the NVEs bosonic mode $n_1=n_2$. The Wigner function for the coherent superposition states $|1\rangle$, $|0\rangle+|1\rangle$, $|0\rangle+|2\rangle$ are shown in Fig. 5. The Wigner function of $|1\rangle$ has one zero-crossing and is radially symmetric, which is as close to a $\delta$ function. The Wigner function for the coherent superposition states $|0\rangle+|1\rangle$ is asymmetric, while the same function for $|0\rangle+|2\rangle$ is centrosymmetric.

In the following, we will demonstrate how to produce some types NVEs entangled states using the method above. For simplicity, we discard the phase factor of the state during the time evolution.

\subsection{NOON State of NVEs}

The maximally entangled $N$ photon state is described as
\begin{eqnarray}\label{15}
|\psi\rangle=\frac{1}{\sqrt{N+1}}\sum_{k=0}^{N}|k,N-k\rangle,
\end{eqnarray}
and the NOON state is in the from \cite{BarryPRA89}
\begin{eqnarray}\label{16}
|\psi\rangle=\frac{1}{\sqrt{2}}(|N_1,0\rangle+|0,N_2\rangle).
\end{eqnarray}
The state with the latter form has an advantage in its sensitivity for optical interferometry over a coherent state and can achieve the Heisenberg limit of $1/N$ in an accuracy of phase measurement.

In order to generate NOON state of two NVEs, the NVEs collective modes are initially prepared in vacuum states, and then we apply a Rabi pulse resonant with the control qubit. When the time is chosen as $\Omega_s t=\pi$, the qubit is prepared in the state $|e\rangle$. Secondly, after the Rabi pulse has been turned off, a controllable shift pulse should be applied via the nonlinear element. In this action, $\Omega$ and $\Delta_d$ are chosen to satisfy $\omega_{z}(t)=\omega_{b1}$. The resonant interaction between NVEs 1 and transmon qubit is fast. Meanwhile, the detuning between the qubit and the NVEs 2 is large ( $G_{2}=0$), so we can consider the NVEs 2 has no effect during the interaction in this step.

After an interaction time $t_{1}^{1}$, the system evolves into the state
\begin{eqnarray}\label{17}
|\psi(t_{1}^{1})\rangle=\cos G_1t_{1}^{1}|e\rangle|0,0\rangle-i\sin
G_1t_{1}^{1}|g\rangle|1,0\rangle.
\end{eqnarray}
Then we apply a different shift pulse to the transmon qubit, which brings a resonantly coupling between the qubit and the NVEs 2 (turn on $G_{2}$), while the detuning between the qubit and the NVEs 1 is large (turn off $G_{1}$). After an interaction time $t_{2}^{1}$, the system evolves into the state
\begin{eqnarray}\label{18}
|\psi(t_{2}^{1})\rangle&=&\cos G_1t_{1}^{1}(\cos
G_2t_{2}^{1}|e\rangle|0,0\rangle\nonumber\\&&-i\sin G_2t_{2}^{1}
|g\rangle|0,1\rangle)-i\sin
G_1t_{1}^{1}|g\rangle|1,0\rangle.\nonumber\\&&
\end{eqnarray}
Choosing the interaction time to satisfy $G_2t_{2}^{1}=\pi/2$, the system state is given by
\begin{eqnarray}\label{19}
|\psi(t_{2}^{1})\rangle&=&-i\cos G_1t_{1}^{1}|g\rangle|0,1\rangle-i\sin
G_1t_{1}^{1}|g\rangle|1,0\rangle.
\end{eqnarray}
After applying a Rabi pulse with the pulse time $\Omega_s t=\pi$, the qubit is prepared in the excited state $|e\rangle$. Turning off the Rabi pulse and adding a shift pulse, the qubit will resonantly interact with the NVEs 1. After an interaction time $t_{1}^{2}$, the system evolves into the state

\begin{eqnarray}\label{20}
|\psi(t_{1}^{2})\rangle &=&
-i[\cos G_1t_{1}^{1}(\cos
G_1t_{1}^{2}|e\rangle|0,1\rangle\nonumber\\&& -i\sin
G_1t_{1}^{2}|g\rangle|1,1\rangle) +\sin G_1t_{1}^{1}(\cos
\sqrt{2}G_1t_{1}^{2}|e\rangle|1,0\rangle\nonumber\\&&
-i\sin\sqrt{2}G_1t_{1}^{2}|g\rangle|2,0\rangle)],
\end{eqnarray}
If the interaction time is chosen as $\cos\sqrt{2}G_1t_{1}^{2}=\sin G_1t_{1}^{2}=0$, the system is on the state
\begin{eqnarray}\label{21}
|\psi(t_{1}^{2})\rangle&=&-i
(\cos G_1t_{1}^{1}|e\rangle|0,1\rangle\nonumber\\&& -i\sin
G_1t_{1}^{1}|g\rangle|2,0\rangle),
\end{eqnarray}
Then let the qubit resonantly interact with the NVEs 2 through another shift pulse, after an interaction time $t_{2}^{2}$, which satisfies $\cos\sqrt{2}G_2t_2^{2}=0$, the system is in the state
\begin{eqnarray}\label{22}
|\psi(t_{2}^{2})\rangle&=&-i[|g\rangle\otimes(\cos G_1t_{1}^{1}|0,2\rangle +\sin
G_1t_{1}^{1}|2,0\rangle)],
\end{eqnarray}
Repeating the process $N$ times, and choosing a suitable interaction time ($\cos\sqrt{N}G_1t_1^{N}=\sin\sqrt{N-1}G_1t_1^{N}=0$ and $\cos\sqrt{N}G_2t_2^{N}=0$), the system state will collapse onto the state
\begin{eqnarray}\label{23}
|\psi(t_{2}^{N})\rangle=-\frac{1}{\sqrt{2}}[|g\rangle\otimes(
|0,n\rangle +|n,0\rangle)],
\end{eqnarray}
here we assume that $\cos G_{1}t_1^{1}=\sin G_{1}t_1^{1}=1/\sqrt{2}$ and discard the phase factor. Then the collective NVEs spins are in the multi-particle NOON state, and the control qubit is in the ground state. From the processes above, NOON state can be produced by transferring amplitude along certain paths in the Fock-state diagram in Fig. 4. This sequence requires a linear number of operations. The corresponding interaction time is given by
\begin{eqnarray}\label{24}
t_{NooN}=(2N-1)\frac{\pi}{\Omega_{s}}+\frac{\pi}{(G_{1}+G_{2})}
\sum_{j=1}^{N}\frac{1}{\sqrt{j}},
\end{eqnarray}

\subsection{Multi-dimensional Entangled States of NVEs}
 The multi-dimensional entangled states are described as
\begin{eqnarray}\label{25}
|\psi\rangle=\frac{1}{\sqrt{N+1}}\sum_{k=0}^{N}|k,k\rangle.
\end{eqnarray}
Compared with low-dimensional entanglement, high-dimensional entanglement, i.e., entangled qudits (the dimension $d\geq 3$), has been proved to be stronger in the violations of local realism \cite{KaszlikowskiPRL00,CollinsPRL02}, and more resilience to error than two-dimensional systems \cite{FujiwaraPRL03}. Besides, quantum cryptographic protocols where qubits are replaced by qudits not only include higher information density coding \cite{DurtPRA04}, but also realize the schemes faster\cite{VaziriPRL02}. Bell inequalities and time-energy degree of freedom for multipartite qudits have been studied \cite{SonPRL06,RichartAPB12}. The application of qudits offers interesting alternatives. For example, they allow the reduction of elementary gates and the number of physical information carriers. Recently, some theoretical schemes have been proposed for
implementing three-dimensional atomic entangled state in cavity QED systems \cite{DelgadoPLA07,YePRA08,LinPRA07,songEPJD08}. For instance, based on quantum Zeno dynamics, Shen et al. \cite{shenJPB11} have proposed a scheme to generate a four-dimensional entangled state between two atoms trapped in two separate uterine cavities.

Now, we will show how to produce high-dimensional entangled states of NVEs based on the physical model above. The two NVEs collective modes are initially prepared in vacuum states and the control qubit is initially prepared in the excited state $|e\rangle$.  Then, a controllable "shift" pulse is applied to satisfy the resonant interaction between the control qubit and NVEs 1, while the NVEs 2 is not affected during the interaction for the large detuning. After an interaction time $t_1^{1}$, $R$ takes effect, i.e. $|e\rangle\rightarrow |g\rangle $ and $|g\rangle\rightarrow |e\rangle $. Then another controllable "shift" pulse is applied to satisfy the resonant interaction between the control qubit and NVEs 2, while the NVEs 1 is large detuning from the control qubit. Choosing an interaction time $t_2^{1}$, the system evolves into the state

\begin{eqnarray}\label{26}
|\psi(t)\rangle&=&\cos
G_{1}t_1^{1}|g,0,0\rangle\nonumber\\&&-i\sin G_{1}t_1^{1}(\cos
G_{2}t_2^{1}|e,1,0\rangle-i\sin G_{2}t_2^{1}|g,1,1\rangle).
\end{eqnarray}
Choosing the interaction time $t_1^{1}$, $t_2^{1}$ to satisfy $\cos G_{1}t_1^{1}=\sqrt{2}/2$ and $\cos G_{2}t_2^{1}=0$, the system state can be reduced to

\begin{eqnarray}\label{27}
|\psi(t)\rangle&=&\frac{\sqrt{2}}{2}[|g\rangle\otimes(|0,0\rangle+|1,1\rangle)].
\end{eqnarray}
The control qubit is in the ground state and the two NVES are in the two-dimensional entangled states.

The second step, we use a Rabi pulse to realize a single qubit rotation, i.e. $|g\rangle\rightarrow|e\rangle$. Implementing the process above once again, after an interaction time $t_1^{2}$ and $t_2^{2}$, the system evolves into the state

\begin{eqnarray}\label{28}
|\psi(t)\rangle&=&\frac{\sqrt{2}}{2}[\cos
G_{1}t_1^{2}|g,0,0\rangle\nonumber\\&&-i\sin G_{1}t_1^{2}(\cos
G_{2}t_2^{2}|e,1,0\rangle-i\sin
G_{2}t_2^{2}|g,1,1\rangle)\nonumber\\&&+\cos
\sqrt{2}G_{1}t_1^{2}(\cos G_{2}t_2^{2}|g,1,1\rangle-i\sin
G_{2}t_2^{2}|e,1,0\rangle)\nonumber\\&&-i\sin
\sqrt{2}G_{1}t_1^{2}(\cos
\sqrt{2}G_{2}t_2^{2}|e,2,1\rangle\nonumber-i\sin
\sqrt{2}G_{2}t_2^{2}|g,2,2\rangle)].
\end{eqnarray}
The interaction time $t_1^{2}$ and $t_2^{2}$ are chosen to be satisfied
$\sin G_{1}t_1^{2}\cos G_{2}t_2^{2}+\cos \sqrt{2}G_{1}t_1^{2}\sin
G_2t_2^{2}=0$ and $\cos \sqrt{2}G_{2}t_2^{2}=0$. Then the system state can be reduced to
\begin{eqnarray}\label{28}
|\psi(t)\rangle&=&|g\rangle\otimes(c_{00}|0,0\rangle+c_{11}|1,1\rangle+c_{22}|2,2\rangle),
\end{eqnarray}
where the control qubit is in the ground state and the two NVEs are in the three-dimensional entangled states. Repeating the process $N$ times, discarding the phase factor and choosing a suitable interaction time each time, the NVEs states will collapse onto the N-dimensional entangled states.

\subsection{Entangled Coherent States of NVEs}

Entangled coherent state was introduced by Sanders \cite{Barry92}. He proposed a scheme of using a nonlinear M-Z interferometer to realize the superpositions of a coherent state and a vacuum state \cite{Barry92A}. Soon the production of entangled coherent state were experimentally realized based on cavity QED system \cite{Davidovich93}. They used one atom traversing two cavities and post-selecting on atomic measurement. Entangled coherent states have been employed in quantum teleportation tasks \cite{Bennett93,Wang01,Johnson02}, which has been used as the entangled resource state employed to affect the teleportation or as the state being teleported.

We consider a transmon qubit coupled to the collective exciton modes of two NVEs with different frequencies. In the strong coupling regime $G\gg \kappa$, where dissipation can be neglected, we can define the Hamiltonian for the system
\begin{eqnarray}\label{29}
H&=&H_0+H_{int},\nonumber\\
H_0&=&\delta_{1}b_1^{+}b_1+\delta_{2}b_2^{+}b_2+\Omega_{s}(t)\sigma_{eg}+\Omega_{s}^{*}(t)\sigma_{ge},\nonumber\\
H_{int}&=&G_{1}(\sigma_{ge}b_1^{+}+\sigma_{eg}b_1)+G_{2}(\sigma_{ge}b_2^{+}+\sigma_{eg}b_2).
\end{eqnarray}
In the interaction picture, the Hamiltonian changes to
\begin{eqnarray}\label{30}
H_{int}^{'}&=&\frac{1}{2}(|+\rangle \langle +|-|-\rangle\langle
-|+e^{i2\Omega_s t}|+\rangle\langle -|+e^{-i2\Omega_s
t}|-\rangle\langle +|)\nonumber\\&& \times(G_{1}b_1e^{-i\delta_1
t}+G_{2}b_2e^{-i\delta_2 t})+H.c.
\end{eqnarray}
Where the dressed states $ \left| \pm \right\rangle=(\left| g \right\rangle\pm\left| e \right\rangle)/\sqrt{2}$ are the eigenstates of $\sigma_{x}=\sigma_{eg}+\sigma_{ge}$ with the eigenvalues $\pm 1$, respectively.  In the strong driving limit $\Omega_s >> \{\delta, G \}$, we can realize a rotating-wave approximation and neglect the terms that oscillate fast. The effective Hamiltonian can be written as \cite{SolanoPRL03}
\begin{eqnarray}\label{31}
H_{eff}^{'}=\frac{\sigma_{x}}{2}(G_1b_1e^{-i\delta_1
t}+G_{2}b_2e^{-i\delta_2 t})+H.c.
\end{eqnarray}
If the two NVEs collective modes are prepared in the vacuum states and the qubit is prepared in the ground state $|g\rangle$. The initial state of the system is
$|g\rangle|0\rangle|0\rangle=(|+\rangle+|-\rangle)|0\rangle|0\rangle/\sqrt{2}$, and
the system at time $t$ will be
\begin{eqnarray}\label{32}
|\psi(t)\rangle=\frac{1}{\sqrt{2}}(|+\rangle|\alpha\rangle|\beta\rangle+|-\rangle|-\alpha\rangle|-\beta\rangle)
\end{eqnarray}
with $\alpha=G_1(e^{i\delta_1 t}-1)/2\delta_1$, $\beta=G_2(e^{i\delta_2 t}-1)/2\delta_2$. Taking a measurement of the control qubit will produce even or odd coherent state of two
NVEs.

\section{Discussion}

Now we investigate the fidelity of the arbitrary state of two NVEs. As an example, the initial state is considered to be $|\Psi(0)\rangle=|e\rangle|0\rangle_{c}(|10\rangle+|10\rangle)_{NVE}/\sqrt{2}$. Turing on $G1$ for $t_{1}=\pi/4G1$, then applying a laser to  realize a state-selective qubit rotation, finally turning on $G2$ for $t_{2}=\pi/4G2$, we get a final system state $|\Psi(t)\rangle$.
The fidelity \cite{UhlmannRMP273} of the prepared states is given by
\begin{eqnarray}\label{33}
F_{total}=\prod_{n_1}\prod_{n_2}F_{n_1,n_2},
\end{eqnarray}
and
\begin{eqnarray}\label{34}
F_{n_1}=(Tr[(\sqrt{\rho_{0}}\rho\sqrt{\rho_{0}})^{1/2}])^2  \quad  t\in t_{1}, \cr
F_{n_2}=(Tr[(\sqrt{\rho_{0}^{'}}\rho^{'}\sqrt{\rho_{0}^{'}})^{1/2}])^2  \quad  t\in t_{2},
\end{eqnarray}
where $\rho_{0}$ ($\rho_{0}^{'}$) and $\rho$ ($\rho^{'}$) correspond to the density matrix of initial and final state in different steps. We plot the fidelity as functions of the frequency of microwave and the detuning between cavity and NVEs, with considering no decay for cavity and collective spin modes. The result of numerical simulation shows that the fidelity keeps being high values $(\geq 0.995)$ when the scaled frequency $\Omega/G$ is evaluated within the domain $[0.25,2]$, as shown in Fig. 6(a). What's more, the optimal fidelity of entangled states is almost unaffected when the fluctuation of the Rabi frequency of the classical field becomes large, which will reduce the difficulty in the experiment. The fidelity is larger than 0.95 when $\Delta/G\geq 60$, and the fidelity keeps being high values $(\geq 0.992)$ when the scaled $\Delta/G\geq 100$, as is shown in Fig. 6(b). In a specific range, the larger the detuning is, the higher the fidelity is, the result show the numerical simulation is in good agreement with the physical model above. The large detuning prolongs the evolution time will lead to the worse impacts of decoherence. It would be interesting to perform a numerical analysis taking into
account the decay of collective spin mode, the spontaneous emission of superconducting qubit, and cavity losses. The master equation of the whole system can be expressed by

\begin{eqnarray}\label{35}
\frac{d\rho}{dt}&=&-i\left[H,\rho \right] +\frac{\kappa}2
(2a\rho a^{+}-a^{+}a\rho -\rho {a}^{+}a)\nonumber\\
&&+\frac{\gamma_{s}}2\sum_{j=1,2}(2b_{j}\rho b_{j}^{+}-b_{j}^{+}b_{j}\rho -\rho {b}_{j}^{+}b_{j})
\nonumber\\&&+\frac{\gamma_{q}}2(2{\sigma
}_{ge}\rho {\sigma }_{eg}-{\sigma }_{eg}{\sigma }_{ge}\rho -\rho
{\sigma }_{eg}{\sigma }_{ge}),
\end{eqnarray}
where $\kappa$, $\gamma_{s}$, and $\gamma_{q}$ denote the effective decay rate of the cavity, collective spin mode and transmon qubit. For simplicity, here, we assume $\gamma _{s}=\gamma _{q}=\gamma$, and consider the same initial state, the same parameters as those in Fig.6. By solving the master equation numerically, we obtain the relation of the fidelity versus the scaled ratio $\gamma/G$ and $\kappa/G$ with $\Delta/G=100$ in Fig. 7.  We see that the collective spin mode decay dominates the reduction of fidelity, while the decay rates of the cavity influence the fidelity slightly, which can be understood by the virtual excitation of the cavity field mode. If we choose $\Delta/G=200$, we can improve the fidelity, while the interaction time needed will be longer.

\begin{figure}[tbp]
 \centering
  \includegraphics[width=0.5\textwidth]{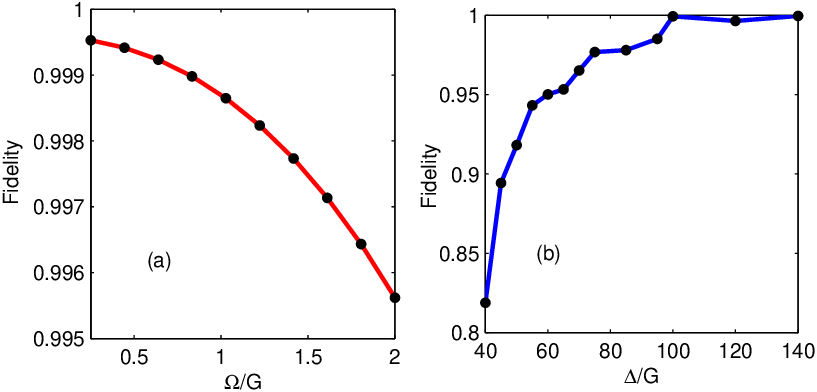} \caption{(Color online)
The fidelity of an arbitrary state of two NVEs versus
the experiment parameters. Here, we consider an initial state $|e\rangle|0\rangle_{c}(|10\rangle+|10\rangle)_{NVE}/\sqrt{2}$,
and no decay for cavity and collective spin modes. The parameters are $G_{1}=G_{2}=G$, $g_m=g_{c}/\sqrt{N}$, $g_c=10G$, $N\sim10^{12}$ and  (a) the scaled Rabi frequency
$0.25\leq\Omega/G\leq 2$ and $\Delta=100G$ (b) the scaled detuning
$40\leq\Delta/G\leq 140$ and $\Omega=G$. }
\end{figure}

\begin{figure}[tbp]
 \centering
  \includegraphics[width=0.5\textwidth]{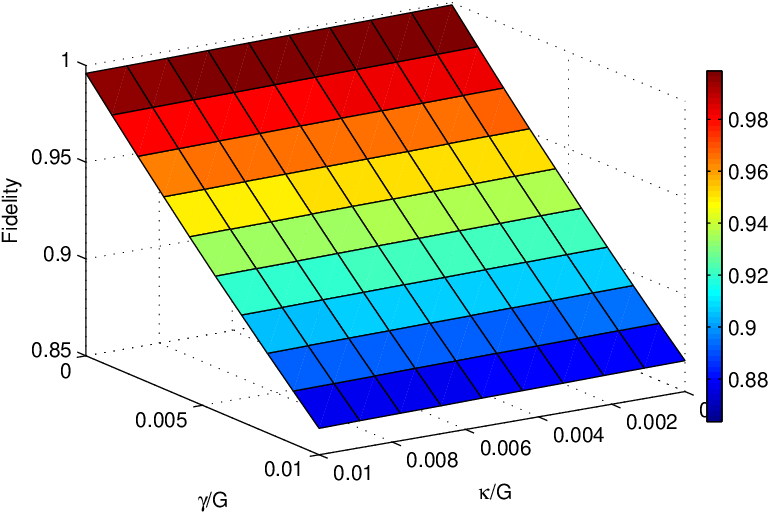} \caption{(Color online)
The fidelity of an arbitrary state of two NVEs versus the scaled decay rates $\kappa/G$ and
$\gamma/G$. Here, we consider an initial state $|e\rangle|0\rangle_{c}(|10\rangle+|10\rangle)_{NVE}/\sqrt{2}$, and the other parameters $\gamma _{s}=\gamma _{q}=\gamma$, $G_{1}=G_{2}=G$, $\Delta/G=100$, $\Omega=G$, $g_m=g_{c}/\sqrt{N}$, $g_c=10G$, $N\sim10^{12}$. }
\end{figure}

In order to bridge the difference in frequency between qubit and NVEs, the SCM cavity frequency $\omega_c$ can be tuned on a nanosecond time scale by applying current pulses through an on-chip flux line, inducing a magnetic flux through a SQUID embedded in the
cavity \cite{PalaciosJLTP08}. $\omega_c$ is varied in order to transfer coherently quantum information between qubit and NVEs. Single qubit rotation and flux pulses placing NVEs and cavity in and out resonance have been realized in a hybrid quantum circuit system \cite{PRL220501}. Qubit state readout is performed by measuring the phase of a microwave
pulse reflected on the nonlinear resonator, which depends on the qubit state; the probability $P_e$ to find the qubit in $|e\rangle$ is then determined by repeating $\sim10^4$ times the same experimental sequence. In this scheme, we have ignored the phase
factors that arise when adding the different shift pulse. These phases can be corrected by including brief pauses between the Rabi and shift pulses \cite{LNature09}, and these do not significantly affect the time evolution during the whole process.

To satisfy the requirement of the physical model, we choose a collective spin coupling strength $\sqrt{N}g_{m}=2\pi\times 10$MHz for an ensemble of $N\sim 10^{12}$ spins, which is consistent with experimentally observed value \cite{PRL220501}. In our case, we choose the microwave and optical detuning $\Delta\sim 2\pi\times 100$MHz, and the Rabi frequency of driving field $\Omega\sim 2\pi\times10$MHz. Then we obtain the effective coupling strength $G=\sqrt{N}g_{m}g_{c}/\Delta\sim 2\pi\times 1$MHz \cite{RevMP623}. We set a decay rate of microwave superconducting cavity $\kappa \sim 2\pi\times 10$kHz \cite{RevMP623} and collective spin decay rate of NVE $\gamma_{s} \sim 2\pi\times 10$kHz \cite{PRB201201}. The inhomogeneous broadening caused by nitrogen electronic spins or a $^{13}$C spin bath maybe affect the desphasing time ($T_{2}$) of NV center. While, we can reduce the impact by narrowing of the nuclear field distribution or the spin-echo techniques, which will prolong the desphasing time from $T_{2}^{*}$ to $T_{2}$ \cite{PRL023603,PRL250503}. By tuning the transmon qubit or the cavity to produce a large detuning of the NVEs transitions or transmon qubit from the cavity, the decoherence of the NVEs-qubit system will be reduced by $\sim 10^{6}$ \cite{APRL09}. The total preparing time of the arbitrary state of two NVEs is associated with the coupling strength $G_i$ in accord to the Eq. (13). Because of
$G=\sqrt{N}g_{m}g_{c}/\Delta$, we emphasize that the growth of the number of NV centers in each spin ensemble could greatly reduce the operation time. In this scheme, we assume $|\Omega_s|<\lambda$ to avoid nonresonant transitions, where $\lambda={G}^2/(\omega_{z}(t)-\omega_{b})$. According to the parameters $\lambda=50G$ and $|\Omega_s|=5G$, the generation of an arbitrary state of two NVEs as the case above will take only $450ns$. The time compares quite favorably to the coherence time of the superconducting qubit, which is now about $10-100\mu s$\cite{RevMP623}. What's more, the implement time is also shorter than the coherence time of NVE \cite{MizuochiPRB09}. So it is possible to efficiently manipulate and measure the entangled states of NVEs in experiments.

\section{conclusion}

In summary, we have shown how to realize the J-C model with collective NVE spin modes. With the help of shift microwave pulses, a tunable qubit alternately resonant interacts with two NVEs, which are coupled to a superconducting cavity. The model provides a possibility for engineering arbitrary entangled states of two distant NVEs. The coupling between the qubit and the NVEs is induced by the non-resonant cavity mode, which is always in the vacuum state. Thus, the evolution of the system is insensitive to cavity decay. The idea can also be used for the preparation of NOON states, N-dimensional entangled states and coherent states of two NVEs. What's more, the scheme can be applied to more NVEs and opens up a way to implement quantum information processing with NVEs-circuit cavity system.

\section{acknowledge}
This work is supported by the National Fundamental Research
Program People's Republic of China under Grant No. 2012CB921601, and the  Research Program of Fujian Education Department under Grant No. JA14044, and the  Research Program of Fuzhou University.


\begin{thebibliography}{ }

\bibitem{RaimondRMP01}J. M. Raimond, M. Brune and S. Haroche, Rev. Mod. Phys.\textbf{ 73}, 565 (2001).

\bibitem{MakhlinRMP01}Y. Makhlin, G. Schon, and A. Shnirman, Rev. Mod. Phys.\textbf{ 73}, 357 (2001).

\bibitem{WallraffNature04} A. Wallraff, D. I. Schuster, A. Blais, \emph{et al.}, Nature (London)\textbf{431}, 162 (2004).

\bibitem{BlaisPRA04} A. Blais, R. S. Huang, A. Wallraff, S. M. Girvin,
         and R. J. Schoelkopf, Phys. Rev. A \textbf{69}, 062320 (2004).

\bibitem{MariantoniNP11} M. Mariantoni, H. Wang, R. C. Bialczak, et.al., Nat.
         Phys.\textbf{7}, 287 (2011).

\bibitem{PRL015502}P.-B. Li, Z.-L. Xiang, Peter Rabl, and Franco Nori, Phys. Rev. Lett. \textbf{117}, 015502 (2016).


\bibitem{YouNature11}J.-Q. You and F. Nori, Nature \textbf{474}, 589 (2011).

\bibitem{APRL09}A. Imamo$\breve{g}$lu, Phys. Rev. Lett.,\textbf{ 102}, 083602 (2009).

\bibitem{PRAPPl044003} P.-B. Li, Y.-C. Liu, S.-Y. Gao, Z.-L. Xiang, Peter Rabl, Y.-F. Xiao and F.-L. Li, Phys. Rev. Appl. \textbf{4}, 044003 (2015).

\bibitem{KuboPRL10}Y. Kubo, F. R. Ong, P. Bertet, et.al., Phys. Rev. Lett.\textbf{ 105},
140502 (2010).



\bibitem{RPRL11}R. Ams$\ddot{u}$ss, Ch. Koller, et. al., Phys. Rev. Lett. \textbf{107}, 060502  (2011).

\bibitem{SchusterPRL10}D. I. Schuster, A. P. Sears, E. Ginossar, et. al., Phys. Rev. Lett. \textbf{105}, 140501 (2010).

\bibitem{LawPRL96}C. K. Law and J. H. Eberly, Phys. Rev.Lett. \textbf{76}, 1055 (1996).

\bibitem{StrauchPRL10}F. W. Strauch, K. Jacobs and R. W. Simmonds, Phys. Rev. Lett. \textbf{105}, 050501 (2010).


\bibitem{YangPRA12}W.-L. Yang, Z.-Q. Yin, Q. Chen, C.-Y. Chen, and M.Feng, Phys. Rev. A \textbf{85}, 022324 (2012).

\bibitem{PRA032342}Bo Li, P.-B. Li, Yuan Zhou, S.-L. Ma and F.-L. Li, Phys. Rev. A \textbf{96}, 032342 (2017).

\bibitem{PRA042306}P.-B. Li, S.-Y. Gao, H.-R. Li, S.-L. Ma and F.-L. Li,
              Phys. Rev. A \textbf{85}, 042306 (2012).

\bibitem{SadowskiarXiv:1303.3757}P. Sadowski, International J. of Quantum Information \textbf{11}, 1350067 (2013).

\bibitem{BertetPRL02}P. Bertet, S. Osnaghi, P. Milman, A. Auffeves, P. Maioli, M. Brune, J. M. Raimond, and S. Haroche, Phys. Rev. Lett.\textbf{ 88}, 143601 (2002).

\bibitem{HofheinzNature08}M. Hofheinz, E. M. Weig, M. Ansmann, R. C. Bialczak, E. Lucero, M. Neeley, A. D. $\acute{O}$Connell, H. Wang, John M. Martinis and A. N. Cleland, Nature (London) \textbf{454}, 310 (2008).


\bibitem{MeunierPRL05} T. Meunier, S. Gleyzes, P. Maioli, A. Auffeves, G. Nogues, M. Brune, J. M. Raimond and S. Haroche, Phys. Rev. Lett.\textbf{ 94}, 010401 (2005).

\bibitem{RauschenbeutelSCi00}A. Rauschenbeutel, G. Nogues, S. Osnaghi, P. Bertet, M. Brune, J. M. Raimond, and S. Haroche, Science \textbf{288}, 2024 (2000).

\bibitem{SPRA08}S.-B. Zheng, Phys. Rev. A \textbf{77}, 045802 (2008).

\bibitem{LloydSCi93}S. Lloyd, Science, \textbf{261}, 1569 (1993).

\bibitem{MalletNP09}F. Mallet, F. R. Ong, A. Palacios-Laloy, F. Nguyen, P. Bertet, D. Vion, and D. Esteve, Nature Phys.\textbf{ 5}, 791 (2009).

\bibitem{PRL220501}Y. Kubo, C. Grezes, A. Dewes, T. Umeda, J. Isoya, H. Sumiya, N. Morishita, H. Abe, S. Onoda, T. Ohshima, V. Jacques, A. Dr$\acute{e}$au,  J. F. Roch, I. Diniz, A. Auffeves, D. Vion, D. Esteve and P. Bertet, Phys. Rev. Lett. \textbf{107}, 220501 (2011).

\bibitem{ZhengPRL00}S.-B. Zheng and G.-C. Guo, Phys. Rev. Lett., \textbf{85}, 2392 (2000)

\bibitem{MarkkuPRL14}M. P. V. Stenberg, Y. R. Sanders and F. K. Wilhelm, Phys. Rev. Lett., \textbf{113}, 210404 (2014).

\bibitem{FPRL11}F. R. Ong, M. Boissonneault, F. Mallet, A. Palacios-Laloy, A. Dewes, A. C. Doherty, A. Blais, P. Bertet, D. Vion and D. Esteve, Phys. Rev. Lett., \textbf{106}, 167002 (2011).

\bibitem{AlexandrePRA04}A. Blais, R.-S. Huang, A. Wallraff, S. M. Girvin, and R. J. Schoelkopf, Phys. Rev. A \textbf{69}, 062320 (2004).


\bibitem{WangPRL09}H. Wang, M. Hofheinz, M. Ansmann and et. al. Phys. Rev. Lett. \textbf{ 103}, 200404 (2009).

\bibitem{BarryPRA89}B. C Sanders, Phys. Rev. A. \textbf{40}, 2417 (1989).

\bibitem{KaszlikowskiPRL00}D. Kaszlikowski, P. Gnaci$\acute{n}$ski, M. $\dot{Z}$ukowski, W. Miklaszewski, and A. Zeilinger, Phys. Rev. Lett. \textbf{85}, 4418 (2000).

\bibitem{CollinsPRL02}D. Collins, N. Gisin, N. Linden, S. Massar, and S. Popescu, Phys.
Rev. Lett. \textbf{88}, 040404 (2002).

\bibitem{FujiwaraPRL03}M. Fujiwara,M. Takeoka, J.Mizuno and M. Sasaki, Phys. Rev. Lett. \textbf{90}, 167906 (2003).

\bibitem{DurtPRA04}T. Durt, D. Kaszlikowski, J. L. Chen and L. C. Kwek, Phys. Rev. A
\textbf{69}, 032313 (2004).

\bibitem{VaziriPRL02}A. Vaziri, G. Weihs, and A. Zeilinger, Phys. Rev. Lett. \textbf{89}, 240401 (2002).


\bibitem{SonPRL06}W. Son, Jinhyoung Lee, and M. S. Kim, Phys. Rev. Lett. \textbf{96}, 060406 (2006).

\bibitem{RichartAPB12}D. Richart, Y. Fischer and H. Weinfurter, Appl Phys B \textbf{106}, 543 (2012).

\bibitem{DelgadoPLA07}A. Delgado, C. Saavedra, and J. C. Retamal, Phys.
Lett. A \textbf{370}, 22 (2007).

\bibitem{YePRA08}S.-Y. Ye, Z.-R. Zhong, and S.-B. Zheng, Phys. Rev. A \textbf{77}, 014303 (2008).

\bibitem{LinPRA07}G.-W. Lin, M.-Y. Ye, L.-B. Chen, Q.-H. Du, and X.-M. Lin, Phys. Rev. A \textbf{76}, 014308 (2007).

\bibitem{songEPJD08}J. Song, Y Xia, H.-S. Song and B Liu, Eur. Phys. J. D \textbf{50}, 91 (2008).

\bibitem{shenJPB11} L.-T. Shen, H.-Z. Wu and R.-X. Chen, J.Phys. B: At. Mol. Opt. Phys. \textbf{44}, 205503 (2011).

\bibitem{Barry92}B. C Sanders, Phys. Rev. A. \textbf{45}, 6811 (1992).

\bibitem{Barry92A}B. C Sanders, Phys. Rev. A.\textbf{ 46}, 2966 (1992).

\bibitem{Davidovich93}L. Davidovich, A. Maali, M. Brune, J. M. Raimond, and S. Haroche, Phys. Rev. Lett. \textbf{71}, 2360 (1993).

\bibitem{Bennett93}C. H. Bennett, G. Brassard, C. Cr$\acute{e}$peau, R. Jozsa, A. Peres, and W. K. Wootters, Phys. Rev. Lett. \textbf{70}, 1895 (1993).

\bibitem{Wang01}X.-G. Wang, Phys. Rev. A \textbf{64}, 022302 (2001).

\bibitem{Johnson02}T. J. Johnson, Stephen D. Bartlett, and Barry C. Sanders, Phys. Rev. A \textbf{66}, 042326 (2002).



\bibitem{SolanoPRL03}E. Solano, G. S. Agarwal and H. Walther, Phys. Rev. Lett.,\textbf{ 90}, 027903  (2003).

\bibitem{UhlmannRMP273}A Uhlmann, Rep. Math. Phys. \textbf{9}, 273 (1976).


\bibitem{PalaciosJLTP08}A. Palacios-Laloy, F. Nguyen, F. Mallet, P. BertetEmail, D. Vion, D. Esteve, J. Low Temp. Phys. \textbf{151}, 1034 (2008).

\bibitem{LNature09}L. DiCarlo, J. M. Chow, J. M. Gambetta, Lev S. Bishop, B. R. Johnson, D. I. Schuster, J. Majer, A. Blais, L. Frunzio, S. M. Girvin and  R. J. Schoelkopf, Nature(London)\textbf{460},240 (2009).

\bibitem{RevMP623}Z.-L Xiang, S. Ashhab, J. Q. You, and Franco Nori,Rev. Mod. Phys. \textbf{85}, 623 (2013).

\bibitem{PRB201201}P. L. Stanwix, L. M. Pham, J. R. Maze, D. Le Sage, T. K. Yeung, P. Cappellaro, P. R. Hemmer, A. Yacoby, M. D. Lukin, and R. L. Walsworth, Phys. Rev. B \textbf{82}, 201201(R) (2010).

\bibitem{PRL023603}L. J. Zou, D. Marcos, S. Diehl, S. Putz, J. Schmiedmayer, J. Majer, and P. Rabl, Phys. Rev. Lett. \textbf{113}, 023603 (2014).




\bibitem{PRL250503}Brian Julsgaard, C$\acute{e}$cile Grezes, Patrice Bertet, and Klaus M${\o}$lmer, Phys. Rev. Lett. \textbf{110}, 250503 (2013).


\bibitem{MizuochiPRB09}N. Mizuochi, P. Neumann, F. Rempp, J. Beck, V. Jacques, P. Siyushev, K. Nakamura, D. J. Twitchen, H. Watanabe, S. Yamasaki, F. Jelezko, and J. Wrachtrup, Phys. Rev. B \textbf{80}, 041201 (2009).

\end{thebibliography}
\end{document}